\documentclass[preprint,review,12pt]{elsarticle}

\usepackage{lineno,hyperref}
\modulolinenumbers[5]

\journal{Materials Today Communications}

\usepackage{graphicx}% Include figure files
\usepackage{color}

\usepackage{amssymb}
\usepackage{amsmath}

\begin{document}
	
\begin{frontmatter}

\title{Unraveling the interlayer and intralayer coupling in two-dimensional layered MoS$_2$ by X-ray absorption spectroscopy and ab initio molecular dynamics simulations}

\author[ISSP]{Inga Pudza}
\ead{inga.pudza@cfi.lu.lv}

\author[ISSP]{Dmitry Bocharov\corref{db}}
\cortext[db]{Corresponding author}
\ead{bocharov@cfi.lu.lv}

\author[ISSP]{Andris Anspoks}
\ead{andris.anspoks@cfi.lu.lv}

\author[PSI]{Matthias Krack}
\ead{matthias.krack@psi.ch}

\author[DESY,ISSP]{Aleksandr Kalinko}
\ead{aleksandr.kalinko@desy.de}

\author[DESY]{Edmund Welter}
\ead{edmund.welter@desy.de}

\author[ISSP]{Alexei Kuzmin\corref{ak}}
\ead{a.kuzmin@cfi.lu.lv}
\cortext[ak]{Corresponding author}

\address[ISSP]{Institute of Solid State Physics, University of Latvia, Kengaraga street 8, LV-1063 Riga, Latvia}

\address[PSI]{Paul Scherrer Institute, Villigen CH-5232, Switzerland}

\address[DESY]{Deutsches Elektronen-Synchrotron DESY, Notkestr. 85, 22607 Hamburg, Germany}

\begin{abstract}
Understanding interlayer and intralayer coupling in two-dimensional layered materials  (2DLMs) has fundamental and technological importance for their large-scale production, engineering heterostructures, and development of flexible and transparent electronics. At the same time, the quantification of weak interlayer interactions in 2DMLs is a challenging task, especially, from the experimental point of view.  Herein, we demonstrate that the use of X-ray absorption spectroscopy in combination with reverse Monte Carlo (RMC) and ab initio molecular dynamics (AIMD) simulations can provide useful information on both interlayer and intralayer coupling in 2DLM 2H$_c$-MoS$_2$.
The analysis of the low-temperature (10-300~K) Mo K-edge extended X-ray absorption fine structure (EXAFS) using RMC simulations allows for obtaining  information on  the means-squared relative displacements $\sigma^2$ for nearest and distant  Mo--S and Mo--Mo atom pairs. This information allowed us further to determine the strength of the interlayer and intralayer interactions in terms of the characteristic Einstein frequencies $\omega_E$ and the effective force constants $\kappa$ for the nearest ten coordination shells around molybdenum. The studied temperature range was extended up to 1200~K employing AIMD simulations which were validated at 300~K using the EXAFS data.
Both RMC and AIMD results provide evidence of the reduction of correlation in thermal motion between distant atoms and suggest strong anisotropy of atom thermal vibrations within the plane of the layers and in the orthogonal direction. 
\end{abstract}

\begin{keyword}
2D layered materials, interlayer coupling, extended X-ray absorption fine structure, reverse Monte Carlo simulations, ab initio molecular dynamics
\end{keyword}

\end{frontmatter}

%\linenumbers

\newpage

\section{Introduction}

The field of 2D layered materials (2DLMs) is a rapidly growing area of material science, offering significant potential for technological advancements \cite{Gupta2015,Sangwan2022}. During the last decades, 2DLMs have generated significant interest  due to their extraordinary electronic, magnetic, optical, catalytic, transport, and tribological properties \cite{Novoselov2016,Kumar2016,Zhan2019,Gutierrez2020,Kaushik2021,Lanza2022,Hoang2022}.
The 2DLM family encompasses a range of substances, including transition-metal dichalcogenides, 2D oxides and hydroxides, boron nitride, and single-element compounds like silicene, phosphorene, germanene, and graphene \cite{Gupta2015,Zhan2019,Shanmugam2022}. The unique structure of 2DLMs is determined by strong in-plane (intralayer) covalent bonding and weak out-of-plane (interlayer) van-der-Waals (vdW) interactions. 

The strength of the interlayer coupling plays a crucial role in the process of exfoliating thin layers and assembling van-der-Waals heterostructures, which is vital for engineering the electronic, optical, and mechanical properties of 2DLMs for their use in devices \cite{Liu2016,Shi2018,Bian2021}.
Therefore, accurate quantification of this weak interlayer coupling has fundamental, theoretical, and technological implications for 2DLMs large-scale production and the ongoing development of flexible and transparent electronics that utilize these materials. For instance, the  insertion of an atomic layer-deposited TiO$_2$ interlayer between exfoliated MoS$_2$ and  an electrode can improve the performance of MoS$_2$ photodetectors \cite{Pak2018}. At the same time, encapsulating MoS$_2$ layers with hexagonal boron nitride in low-power transistors offers protection from environmental factors and ensures stability in device operation, even at elevated temperatures \cite{Lee2015}. 
Furthermore, future integrated circuits will utilize transistors based on 2D semiconductors that are coupled through  van-der-Waals interactions with high-$\kappa$ dielectric and metal contacts \cite{Li2019,Das2021}. A microprocessor based on an active channel material made of molybdenum disulfide was demonstrated \cite{Desai2016,Wachter2017} and opens up new opportunities for next-generation low-power electronics \cite{Migliato2020logic}.
	
Although the physical and chemical properties of 2DLMs associated with the intralayer bonding have been studied both experimentally and theoretically \cite{Gupta2015,Liu2016,Shi2018}, the extensive atomic and nanoscale characteristics of their interlayer vdW-dependent properties (for example, enhanced in-plane stiffness, band gap opening, band structure, and carrier mobility engineering,  interlayer charge transfer/distribution, etc.) is still a major challenge due to the highly anisotropic nature of the weak interlayer interactions, the lack of accurate experimental methods to quantify such complex interlayer behavior, and complexity of manipulations with ultra-thin and highly transparent 2DLMs.
Therefore,  gaining a deeper insight into the weak interlayer interactions in 2D layered materials and their interactions with different substrates is critical for improving the transfer efficiency and uniform thickness of printed flakes. This, in turn, is important for the production of high-quality, large-scale 2DLM-based devices at the micro- and nanoscale \cite{Ferrari2015,Li2017,Butanovs2018,Polyakov2016}. 

Until now, the experimental approaches for detecting interlayer coupling in 2DLMs  have been limited to photoluminescence, Raman, and angle-resolved photoemission spectroscopy  \cite{Shi2018}.
However, these methods do not give direct information on the atomic structure of the material \cite{Billinge2007}. 
Note also that Raman spectroscopy is a selective method providing information about the frequency of vibrations but not their amplitude. 
Theoretical modeling of vdW interactions is also a difficult task because of its highly non-local nature, so this contribution is often accounted for a \textit{posteriori} \cite{Stohr2019}.  

The present study is largely based on the use of X-ray absorption spectroscopy (XAS) which is a direct local structural tool complementary to diffraction. The technique has been employed in the past to probe the short-range order in 2DLMs  \cite{Chung2014,Caramazza2016,Kuznetsov2019}. However, the conventional approach to the data analysis of the X-ray absorption spectrum, based on multi-shell modeling, cannot provide reliable information on the medium-range (across the vdW spacing) interlayer interactions due to a strong correlation between model parameters for distant coordination shells during fitting \cite{Kuzmin2014,Kuzmin2020}. At the same time, the extraction of such information should be in principle possible since the range of atomic structure around the absorbing atoms probed by XAS  extends often up to 10--15~\AA\ as follows from the calculations of the photoelectron mean-free path \cite{Kuzmin2014,Kuzmin2020}. Nevertheless, this task is challenging and requires the combined use of advanced theoretical methods and high-quality experimental data, i.e. the extended X-ray absorption fine structure (EXAFS) spectra, which can be recorded nowadays at last-generation synchrotron radiation sources.        
Recent developments in the EXAFS data analysis include two approaches based on atomistic simulations -- reverse Monte Carlo (RMC) and molecular dynamics (MD) methods \cite{Kuzmin2020}. 
Both approaches enable precise examination of the local atomic structure and lattice dynamics with high accuracy taking into account multiple-scattering contributions \cite{Rehr2000} and are well suited for the analysis of distant coordination shells, 
for example, interactions between atoms located in neighbouring layers of 2DLMs.
Note that a number of previous works dedicated to the EXAFS studies of MoS$_2$ utilized conventional approach, i.e., the multi-shell fitting, and were limited to the analysis of the first two coordination shells of molybdenum atoms only  \cite{Chung2014,Caramazza2016,Joensen1987,BELYAKOVA2004,Lassalle2015,Zhang2016,KIM2017}.

In this study, a complex approach based on X-ray absorption spectroscopy, ab initio molecular dynamics (AIMD), and reverse Monte Carlo methods was employed to provide insights
into the  local dynamics in 2DLM 2H$_c$-MoS$_2$.  We demonstrated, for the first time, the possibility of extracting the temperature dependence (10-300~K) of intralayer and interlayer interactions in 2DLM 2H$_c$-MoS$_2$ for the nearest ten coordination shells around molybdenum atoms.  This information was obtained through the analysis of the Mo K-edge EXAFS spectra using the reverse Monte Carlo method.  The low-temperature experimental results were complemented with high-temperature (300-1200~K) AIMD simulations, which were validated at 300~K using EXAFS data. Both approaches allowed us to observe similar trends in the lattice dynamics of 2H$_c$-MoS$_2$ and to distinguish between interlayer and intralayer interactions.

\begin{figure}
	\centering	
	\includegraphics[width=0.7\textwidth]{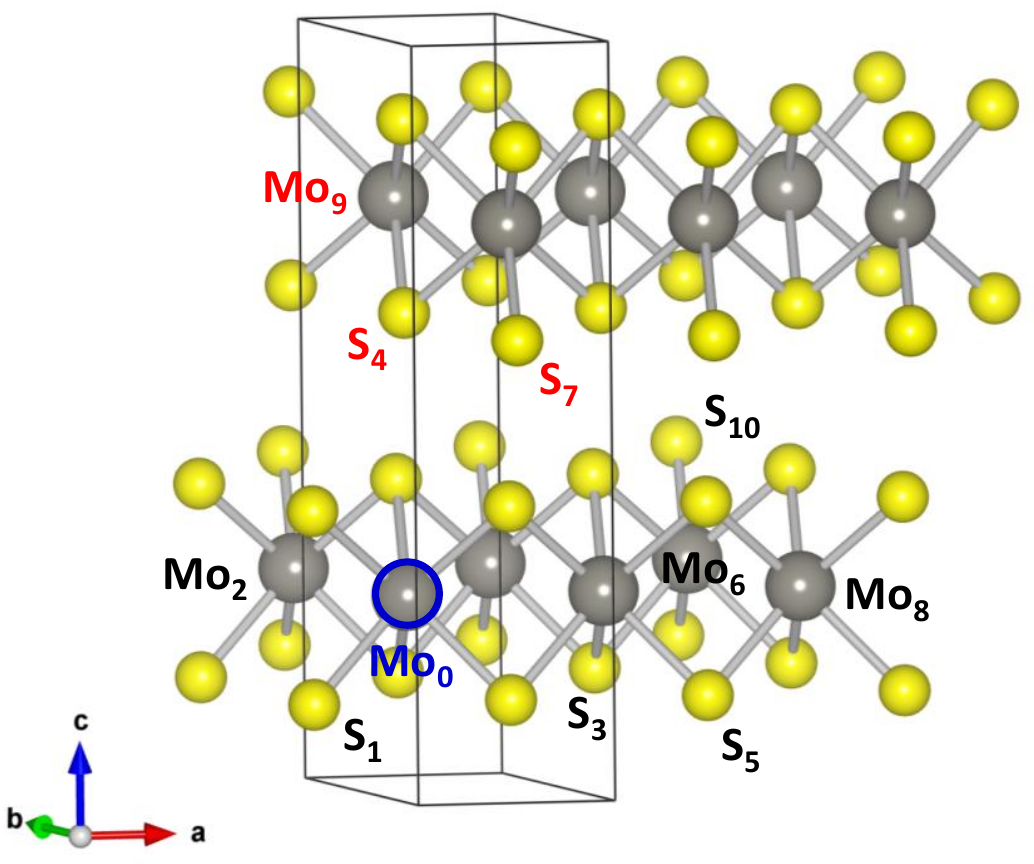}
	\caption{Crystallographic structure of 2H$_c$-MoS$_2$. Atoms located in the nearest ten coordination shells around the Mo$_0$ atom are indicated.}
	\label{fig1}
\end{figure}
  
\section{Experimental}

Commercial molybdenum(IV) sulfide  (MoS$_2$) powder (Sigma-Aldrich, 98\%) was used in all experiments. 
The phase of the sample was confirmed using X-ray powder diffraction and Raman spectroscopy.

The X-ray absorption experiments were conducted at the Mo K-edge as a function of temperature from 10 to 300~K at the DESY PETRA-III P65 beamline \cite{P65}. 
The storage ring was operating at $E$=6.08 GeV and current $I$=100 mA in top-up 480 bunch mode.  A fixed exit  double-crystal Si(311) monochromator was used, and the harmonic rejection was achieved by the Rh-coated silicon plane mirror. 
The X-ray absorption spectra were recorded in transmission mode using two ionization chambers, while the temperature was controlled using a liquid helium flow cryostat.

The sample was prepared by mixing MoS$_2$ powder with cellulose in an agate mortar and pressing the mixture into a pellet using a hydraulic press. Its thickness was optimized to get the value of the absorption edge jump $\Delta \mu$=1. The Mo K-edge EXAFS spectra were extracted following the conventional procedure \cite{Kuzmin2014}.
The Fourier transforms (FTs) of the EXAFS spectra were calculated using 10\% Gaussian function. Note that the peak positions in all FTs differ from their true crystallographic values
because of the EXAFS phase shifts.

\section{Reverse Monte-Carlo simulations}

The structural information encoded in the experimental Mo K-edge EXAFS spectra was extracted using the reverse Monte-Carlo method based on an evolutionary algorithm approach (RMC/EA)   \cite{Timoshenko2012rmc,Timoshenko2014rmc}. 
The simulations were performed by the EvAX code \cite{Timoshenko2014rmc}.
The RMC/EA method involves randomly changing the atomic coordinates within a three-dimensional structure model of the material in order to minimize the difference between the experimental and calculated configuration-averaged EXAFS spectra. Both structural and thermal disorder in the material, as well as multiple-scattering effects, were taken into account.

The initial structural models for the RMC/EA calculations were constructed based on diffraction data (2H$_c$-MoS$_2$, space group $P6_3/mmc$ (194), $a$=$b$=3.160~\AA, $c$=12.295~\AA\ \cite{Wyckoff1963}) in the form of the supercell (4$a$$\times$4$b$$\times$4$c$) containing 384 atoms. The atoms in the supercell were randomly displaced at each iteration with the maximum allowed displacement of 0.4~\AA. 
The configuration-averaged EXAFS spectra at the  Mo K-edge were calculated using the \textit{ab initio} self-consistent real-space multiple-scattering (MS) FEFF8.5L code \cite{Rehr2000,Ankudinov1998} taking into account multiple-scattering contributions up to the 5$^{th}$ order. The complex energy-dependent exchange-correlation Hedin-Lundqvist potential was employed to account for inelastic effects \cite{Hedin1971}. The amplitude scaling parameter $S_0^2$ was set to 1. 

The experimental and calculated Mo K-edge EXAFS spectra were compared in both the direct ($R$) and reciprocal ($k$) space utilizing the Morlet wavelet transforms \cite{Timoshenko2009wavelet}. The fitting was done in the $k$-space range from 2.2~\AA$^{-1}$ to 18.3~\AA$^{-1}$ and the $R$-space range from 1.1~\AA\ to 6.3~\AA. 
The convergence of each RMC simulation was achieved after  4000 iterations. 
The RMC/EA calculations were performed for 32 atomic configurations simultaneously. 
At least three RMC/EA simulations were conducted for each experimental data set, using different sequences of pseudo-random numbers.
The configuration-averaged EXAFS spectra agree well with the experimental data at all temperatures supporting the reliability of the obtained structural models.  

Atomic coordinates were used to calculate radial distribution functions (RDFs) and estimate structural parameters of interest.  The means-square relative displacement factors $\sigma^2$ for Mo--S and Mo--Mo atom pairs were calculated using the median absolute deviation (MAD) method \cite{Hampel1973robust,Daszykowski2007robust}.

\section{Ab initio molecular dynamics}

Ab initio molecular dynamics (AIMD) simulations were based on Kohn-Sham density functional theory (DFT) \cite{Kohn1965} and were performed in the NVT ensemble at four different temperatures (300~K, 600~K, 900~K, and 1200~K) using the CP2K
code \cite{CP2K2020}. The code employs a localized basis set of Gaussian-type orbital functions for the description of the Kohn–Sham matrix within the framework of the Gaussian Plane Waves method \cite{Quickstep,GPW}. We used PBE exchange-correlation functional \cite{Perdew1996} with Grimme correction \cite{Grimme2010}.  

Preliminary structure DFT calculations with full optimization of geometry gave the lattice constants of bulk 2H$_c$-MoS$_2$ equal to $a_0 = 3.164$~\AA\ and $c_0 = 12.199$~\AA.  
Orthorhombic supercell 6$a_0 \times$4$a_0 \times$2$c_0$ with 576 atoms (192 Mo atoms and 384 S atoms) was used for AIMD calculations.  The size of the supercell was equal to 18.986~\AA$\times$21.923~\AA$\times$24.398~\AA.  CSVR (Canonical Sampling through Velocity Rescaling) thermostat \cite{Bussi2007} was used in the calculations. The system was first equilibrated for 15~ps, followed by a production run of 30~ps.
The time step was $\Delta t = 0.5$~fs. A set of at least 4000 atomic configurations was recorded during the production run and further used for the calculation of the configuration-averaged EXAFS spectrum by the MD-EXAFS approach \cite{Kuzmin2014,Kuzmin2009}. The AIMD results calculated at different temperatures were used to obtain temperature-dependence of the means-square relative displacement factors $\sigma^2$ for required Mo--S and Mo--Mo interatomic distances.

\section{Results and discussion}

The crystallographic structure of 2DLM 2H$_c$-MoS$_2$ with hexagonal symmetry is shown in  Figure\ \ref{fig1} \cite{Schonfeld1983}. Molybdenum atoms have trigonal prismatic coordination and are covalently bonded to six sulfur ions. The MoS$_6$ units form two layers per unit cell which are held together along the $c$-axis by weak vdW interactions. The interlayer gap is about 3~\AA. In Figure\ \ref{fig1}, the Mo and S atoms located in the nearest ten coordination shells around the absorbing molybdenum atom (Mo$_0$) are labeled. Note that three shells including sulfur (two S$_4$ and twelve S$_7$) and molybdenum (six Mo$_9$) atoms belong to the nearest neighboring layers, therefore, their analysis can provide original information on the interlayer interactions.  

\begin{figure}
	\centering	
	\includegraphics[width=0.7\textwidth]{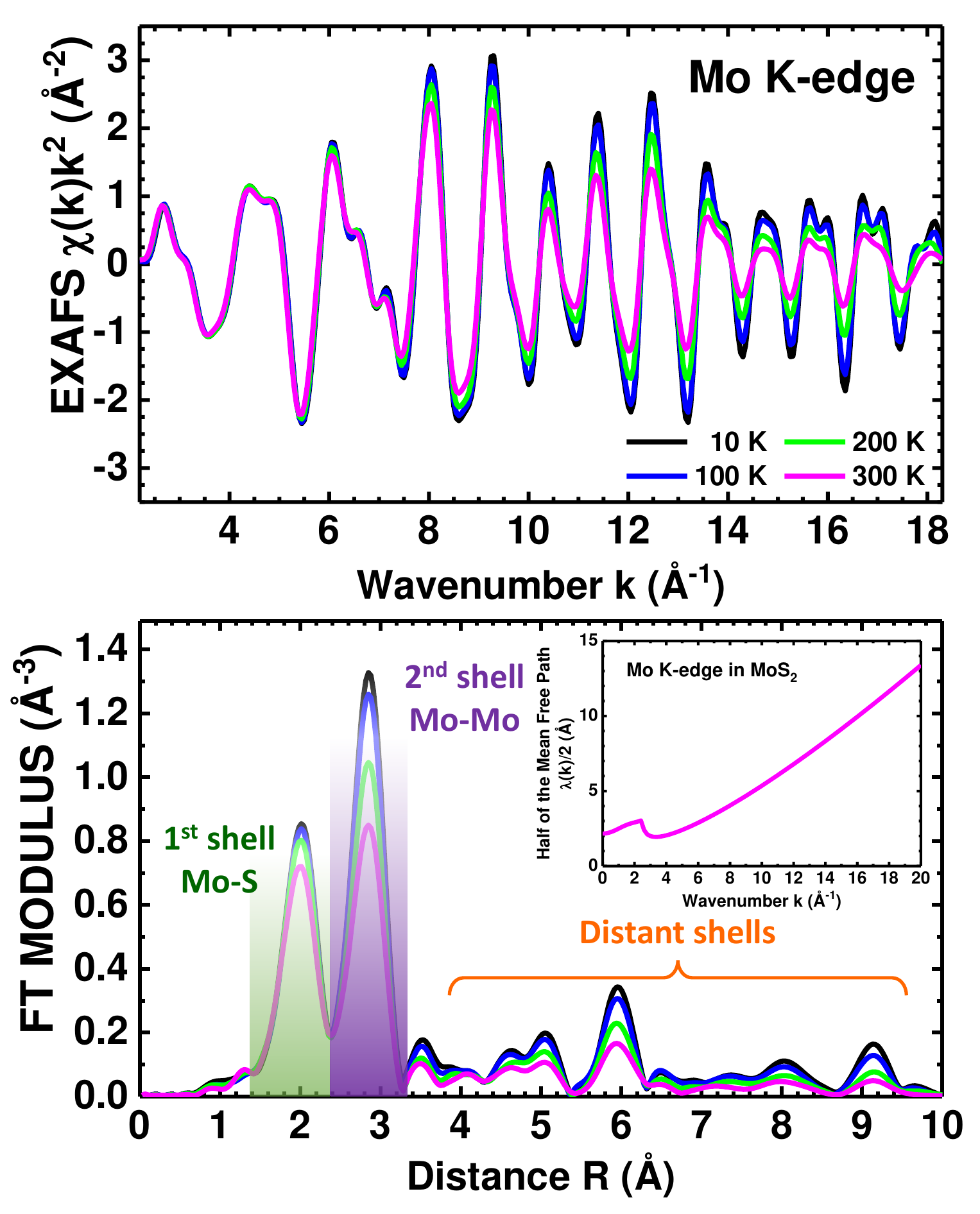}
	\caption{Temperature-dependent Mo K-edge EXAFS spectra $\chi(k)k^2$ and their Fourier transforms for 2H$_c$-MoS$_2$. The calculated half of the mean free path of the photoelectron is shown in the inset and explains the presence of distant coordination shells. }
	\label{fig2}
\end{figure}

The values of Mo--S and Mo--Mo interatomic distances  for the first ten coordination shells
in 2H$_c$-MoS$_2$ are reported in  Table\ \ref{table1}. As can be seen, the atoms (S$_4$, S$_7$, and Mo$_9$) located in the nearest neighboring layers and being the closest to the absorbing molybdenum  (Mo$_0$)  belong to its distant coordination shells with large interatomic distances.
Nevertheless, these atoms should contribute to EXAFS spectra as can be estimated from the value of the mean-free path of the excited 1s(Mo) photoelectron (see, the inset in Figure\ \ref{fig2}). 

The Mo K-edge EXAFS spectra of 2H$_c$-MoS$_2$ and their FTs are dominated at all studied temperatures in the range of 10-300~K  by a contribution from the first two coordination shells, which are composed of six sulfur (S$_2$)  and  six molybdenum (Mo$_2$) atoms  (Figure\ \ref{fig2}).  However, several structural peaks are well visible in FTs above 3.2~\AA\ up to at least 10~\AA. Their small amplitude compared to that of the first two peaks at 2.0 and 2.9~\AA\ is caused by  the overlap of shells, thermal disorder, and interference effects between single-scattering and high-order multiple-scattering contributions in EXAFS spectra. 
In particular, the effect of thermal disorder is clearly seen upon increasing temperature as the EXAFS spectra damping at high-$k$ values or a reduction of the peak amplitude in FTs.  
While the reliable analysis of distant shells is challenging within the conventional methodology, it can be reliably performed using the reverse Monte Carlo method \cite{Timoshenko2014rmc,Jonane2018,Bakradze2021,Kotomin2022}.

\begin{table}[h]
	\centering 
	\caption{Values of Mo--S and Mo--Mo interatomic distances for the first ten coordination shells of molybdenum calculated from the crystallographic structure of 2H$_c$-MoS$_2$ \protect\cite{Schonfeld1983}. Interlayer distances are given in bold. The values of characteristic Einstein frequencies ($\omega_E$) and the effective force constants ($\kappa$) obtained from the RMC analysis are also given.} 
	\label{table1}
	\begin{tabular}{llll}
		\hline
		Atom pair  & Distance (\AA) & $\omega_E$ (THz) & $\kappa$ (N/m)\\
		\hline
		Mo$_0$--S$_1$  & 2.37  &  62$\pm$3	& 152$\pm$13  \\
		Mo$_0$--Mo$_2$  & 3.16  & 44$\pm$2	& 152$\pm$13		 \\
		Mo$_0$--S$_3$  & 3.95   & 38$\pm$2 & 56$\pm$5  		\\
		\textbf{Mo$_0$--S$_4$}  & \textbf{4.64}  & \textbf{26$\pm$2} & \textbf{26$\pm$5}  \\
		Mo$_0$--S$_5$  & 5.06   &  40$\pm$2 & 65$\pm$5 		 \\
		Mo$_0$--Mo$_6$  & 5.48  &  34$\pm$2 & 94$\pm$5 	  \\
		\textbf{Mo$_0$--S$_7$}  & \textbf{5.62}  & \textbf{36$\pm$2} & \textbf{51$\pm$5}   \\
		Mo$_0$--Mo$_8$  & 6.32   &  34$\pm$2 &	93$\pm$6 		   \\		
		\textbf{Mo$_0$--Mo$_9$}  & \textbf{6.41}  &  \textbf{18$\pm$2} & \textbf{25$\pm$5}	 \\
		Mo$_0$--S$_{10}$  & 6.75  &  37$\pm$2 &	55$\pm$5 		    \\
		\hline
	\end{tabular}
\end{table}

\begin{figure}
	\centering
	\includegraphics[width=0.9\textwidth]{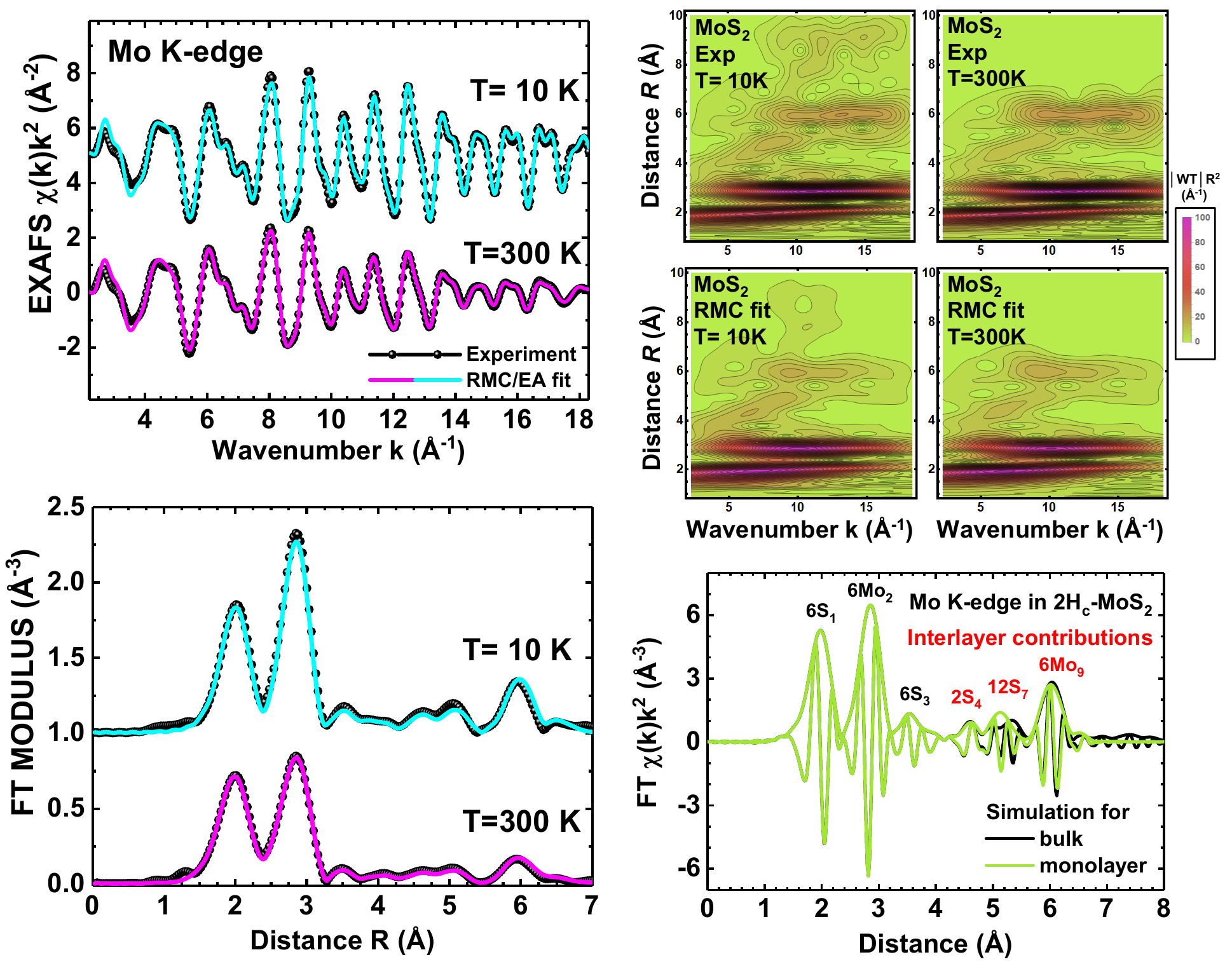}
	\caption{Results of RMC calculations for the Mo K-edge in 2H$_c$-MoS$_2$ at 10~K and 300~K. The EXAFS spectra $\chi(k)k^2$ and their Fourier transforms are shown in the left panels. Wavelet  transforms of EXAFS spectra are shown in the upper right panel. A comparison between the calculated Fourier transforms for the bulk and single monolayer MoS$_2$ is shown in the lower right panel. }
	\label{fig3}
\end{figure}

Examples of the RMC simulations of the Mo K-edge EXAFS spectra for 2H$_c$-MoS$_2$ at two selected temperatures (10 and 300~K) are shown in Figure\ \ref{fig3} in $k$-space, $R$-space, and wavelet($k,R$)-space. The theory reproduces well the experimental data, allowing for a detailed analysis of thermal disorder effects within the coordination shells up to 7~\AA.

To estimate the contribution of atoms from the nearest neighboring layers to the total EXAFS spectrum,
the Mo K-edge EXAFS spectra of bulk and single monolayer MoS$_2$ were additionally calculated using the crystallographic structure of 2H$_c$-MoS$_2$ \cite{Schonfeld1983}, without considering thermal disorder. Their FTs are compared in the lower right panel in Figure\ \ref{fig3}.
As one can see, the main difference between the two FTs is observed between 4.2 and 6.5~\AA, where the interlayer contributions appear. At the same time, their effect is rather small and requires a delicate treatment during analysis.  

\begin{figure}
	\centering	
	\includegraphics[width=0.9\textwidth]{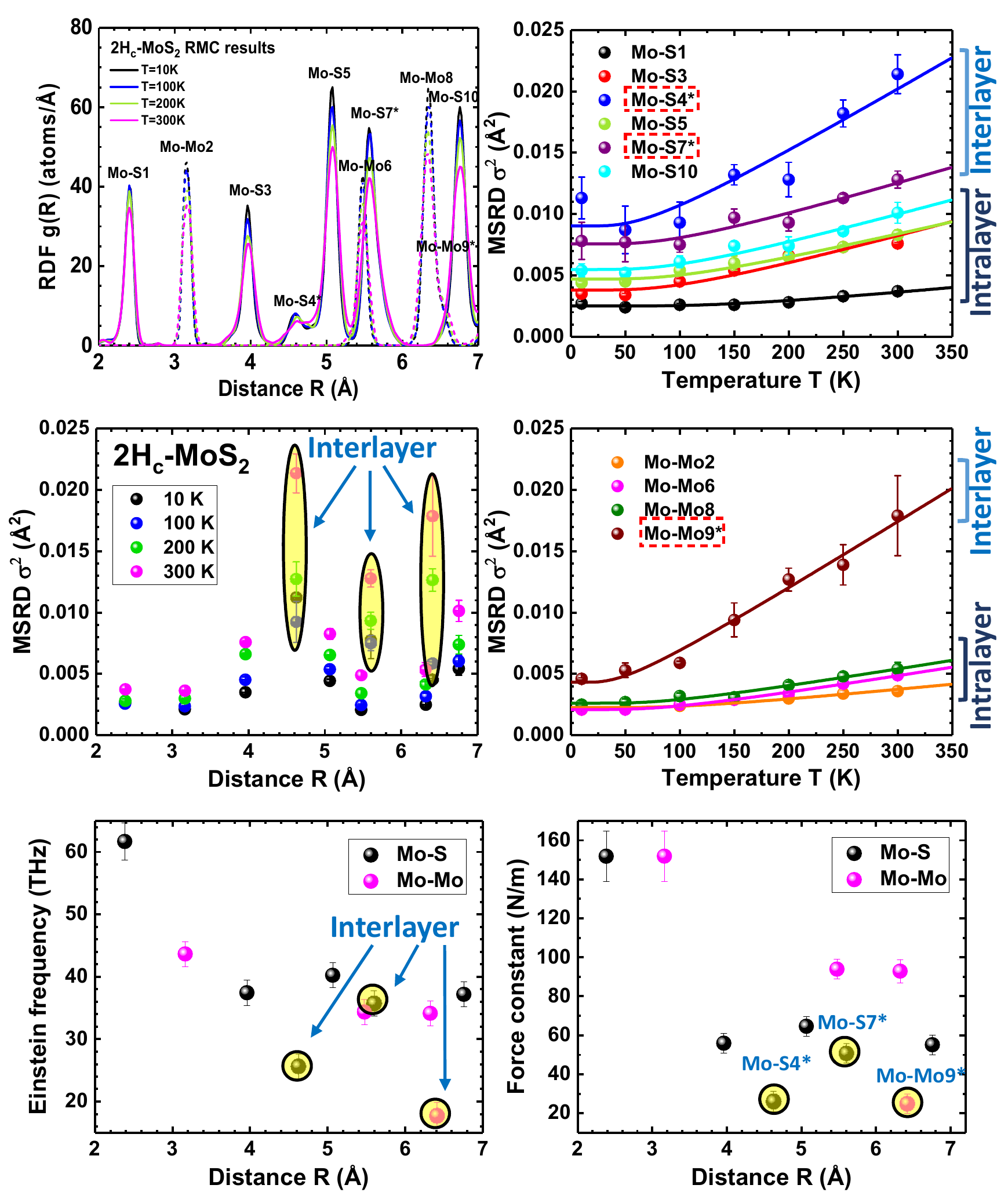}
	\caption{Results of RMC calculations for the Mo K-edge in 2H$_c$-MoS$_2$. Upper left panel: the radial distribution functions for Mo--S and Mo--Mo atom pairs at four temperatures (10~K, 100~K, 200~K, 300~K). Middle left panel: MSRD factors $\sigma^2$($R$) for Mo--S and Mo--Mo atom pairs as a function of interatomic distance for four temperatures. Upper and middle right panels: temperature dependence of the MSRD factors $\sigma^2$($T$) for Mo--S and Mo--Mo atom pairs.  Lower left and right panels: distance dependence of the characteristic Einstein frequencies $\omega_E$ and the effective force constants $\kappa$ for Mo--S and Mo--Mo atom pairs. }
	\label{fig4}
\end{figure}

The coordinates of atoms obtained in the RMC simulations were used to compute the radial distribution functions (RDFs) $g(R)$ and  the means-square relative displacements (MSRDs) $\sigma^2$ at each temperature (Figure\ \ref{fig3}). The temperature dependencies (10-300~K) of the obtained MSRDs  $\sigma^2(T)$ were further  approximated by the correlated Einstein model \cite{Sevillano1979}. As a result, the characteristic Einstein frequencies $\omega_E$  and the effective force constants $\kappa$ were obtained for all coordination shells and are reported in Table\ \ref{table1}\ and Figure\ \ref{fig4}. 
The relationship between the two parameters is $\omega_E = \sqrt{\kappa / \mu}$, where $\mu$ is the reduced mass of an atom pair.
The largest value of the Einstein frequency, $\omega_E$ = 62~THz, was found for the Mo$_0$--S$_1$ bonds in the first coordination shell due to strong covalent bonding between the nearest Mo and S atoms. The effective force constant $\kappa$ = 152$\pm$13~N/m for the Mo$_0$--S$_1$ bonds can be compared to the one (194~N/m) estimated from the internal Raman E$^1_{2g}$ mode at 383.4~cm$^{-1}$ \cite{Bagnall1980} or to the one (138~N/m) determined from the analysis of dispersion-curves measured by neutron scattering \cite{Wakabayashi1975}.  In the EXAFS study \cite{Caramazza2016}, the effective force constant $\kappa$ = 204~N/m  for the Mo$_0$--S$_1$ bond was obtained by the conventional EXAFS analysis.

The characteristic Einstein frequency $\omega_E$ = 44~THz is lower for the Mo$_0$--Mo$_2$ atom pairs due to the twice larger value of the respective reduced mass ($\mu$(Mo--O)=24.03 vs $\mu$(Mo--Mo)=47.97). Nevertheless, the values of the effective force constants $\kappa$ = 152~N/m are equal for the two atom pairs. This fact can be explained by the location of each molybdenum atom at the center of a trigonal prism composed of six sulfur atoms, while the prisms themselves are connected by edges (Figure\ \ref{fig1}). Thus, while there is no direct bonding between neighbouring molybdenum atoms, they  interact through the two common sulfur atoms at the prism edge. Note that the value of  $\kappa$ = 137~N/m was found for the Mo$_0$--Mo$_2$ atom pair in \cite{Caramazza2016}. 

For atom pairs involving distant atoms (S$_4$, S$_7$, and Mo$_9$) from the nearest neighboring layers, the characteristic Einstein frequencies and the effective force constants have the lowest values suggesting weak interlayer coupling.

The analysis of MSRD factors (Figure\ \ref{fig4}) indicates significant differences  in relative vibrations between  atoms in the same layer and  the nearest neighboring layers.  
In general, a linear increase in MSRD values for Mo--S and Mo--Mo atom pairs is expected at higher temperatures due to stronger thermal vibrations \cite{Sevillano1979}. 
As the separation between two atoms grows, the MSRD value for the pair of atoms should eventually reach a limiting value equal to the sum of their mean-squared displacements, owing to a decrease in the correlation of their motion  \cite{Jeong2003}.
Indeed, such behavior was observed in all coordination shells except the ones located in the
nearest neighboring layers. The MSRD factors for Mo$_0$--S$_4$, Mo$_0$--S$_7$, and Mo$_0$--Mo$_9$ atom pairs are significantly  larger  than those for other atom pairs at all temperatures.  This result suggests that a weak vdW coupling between layers in 2H$_c$-MoS$_2$ leads to highly anisotropic vibrations of Mo and S atoms, being smaller within the layers and larger along the $c$-axis. To summarize, the obtained results demonstrate the sensitivity of EXAFS spectra to such a weak effect as an interlayer coupling in 2DLMs. 

\begin{figure}
	\centering	
	\includegraphics[width=0.9\textwidth]{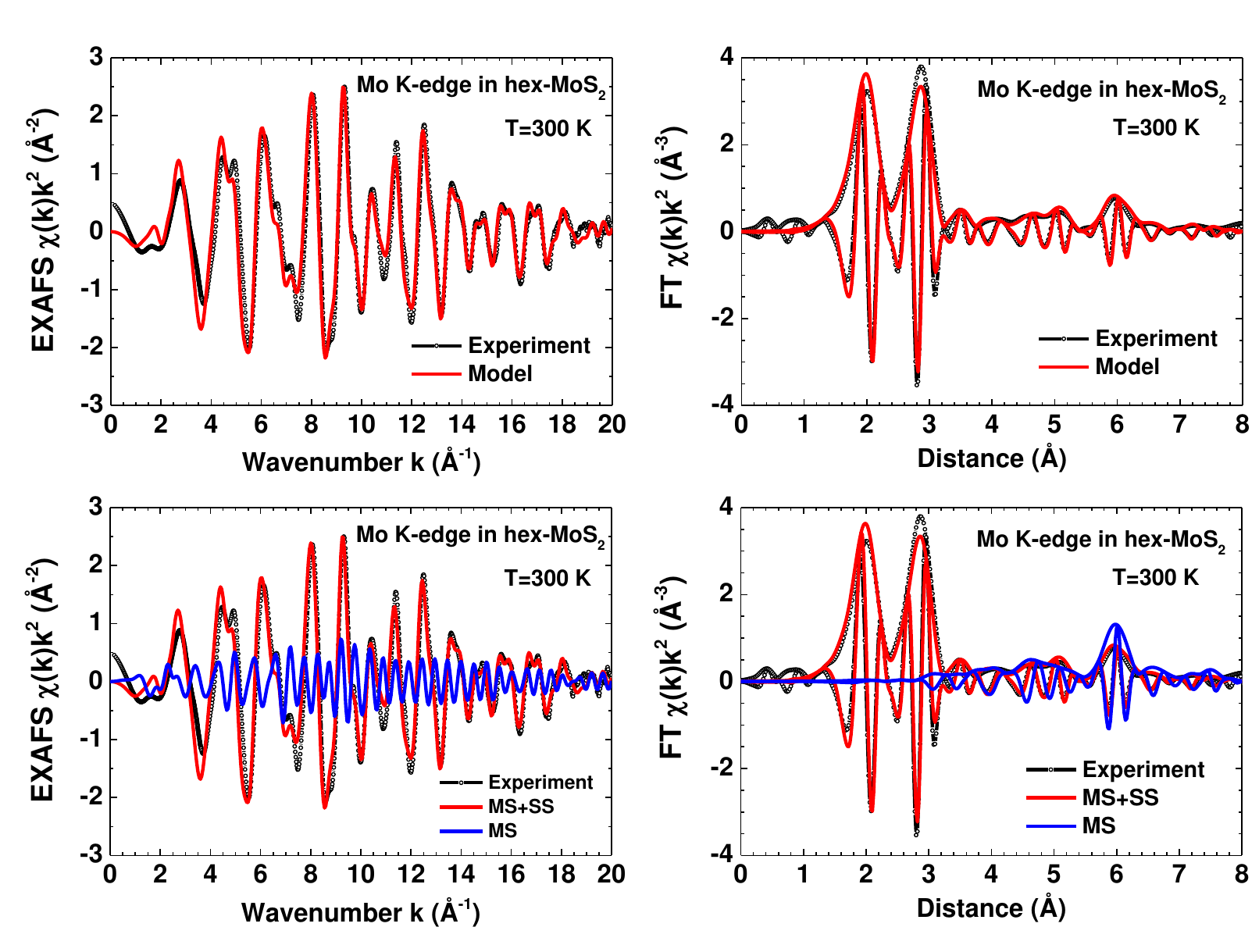}
	\caption{Comparison between the experimental and AIMD calculated Mo K-edge EXAFS spectra $\chi(k)k^2$ and their Fourier transforms for 2H$_c$-MoS$_2$. The single-scattering and multiple-scattering  contributions to the total EXAFS spectrum are shown in the lower panels.  }
	\label{fig5}
\end{figure}

To support experimental findings and extend the analysis of lattice dynamics in 2H$_c$-MoS$_2$ to even higher temperatures, ab initio molecular dynamics (AIMD) simulations were performed. The Mo K-edge EXAFS spectrum at 300~K was used to validate the AIMD theory. A comparison between the experimental and AIMD-calculated EXAFS spectra is shown in Figure\ \ref{fig5} and suggests good agreement both in $k$ and $R$ spaces. 
The AIMD simulations reproduce well the structure and lattice dynamics in  2H$_c$-MoS$_2$ at least up to 7~\AA.
Note also that the range above the second shell is strongly dominated by the multiple-scattering contributions (blue curves in Figure\ \ref{fig5}) which, thus, should not be ignored and  must be taken into account for accurate analysis.

\begin{figure}
	\centering	
	\includegraphics[width=0.9\textwidth]{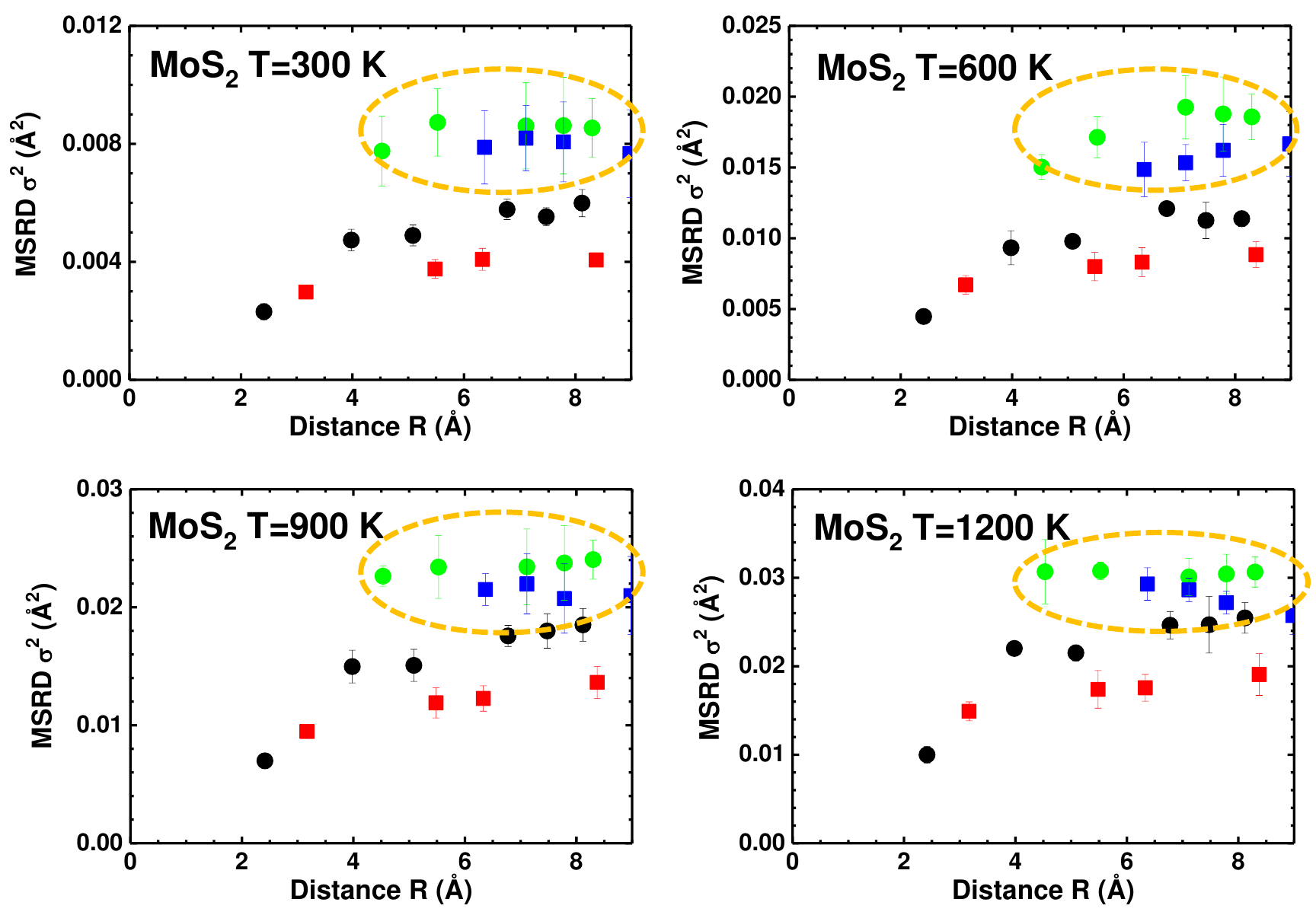}
	\caption{Temperature and interatomic distance dependences of the MSRD factors $\sigma^2$($T$) for Mo--S (circles) and Mo--Mo (squares) atom pairs obtained from AIMD simulations for 2H$_c$-MoS$_2$. The MSRD factors for interlayer interactions are indicated by dashed ellipses.}
	\label{fig6}
\end{figure}

Being inspired by the results at 300~K, the AIMD simulations were also performed at higher temperatures of 600~K, 900~K, and 1200~K. The MSRD factors for Mo--S and Mo--Mo atom pairs were calculated from atomic coordinates after system equilibration and are shown in Figure\ \ref{fig6} for different coordination shells at four temperatures. Note that the MSRD factors for Mo--S atom pairs are slightly larger than those for Mo--Mo atom pairs when close interatomic distances are considered. 
The results obtained allow us to draw several conclusions. Firstly,  upon temperature growth, the absolute values of the MSRD factors increase due to the larger amplitude of atom thermal vibrations. Secondly, at each temperature, the MSRD factors increase up to about 4~\AA\ due to a decrease in the correlation of atomic motion, i.e. due to a reduction of interaction between distant atoms. At the same time,  the MSRD values of Mo--S and Mo--Mo atom pairs remain almost unchanged for more distant coordination shells (beyond 4~\AA) where the correlation is absent, and the MSRD factors equal to a sum of mean-squared displacements for two atoms. 
Thirdly,  there is a clear difference between the MSRD values for atom pairs within the layer and including atoms from the nearest neighboring layers indicated in Figure\ \ref{fig6} by dashed ellipses.  
The origin of this difference is weak atomic interactions across the vdW gap leading to the  anisotropic atomic motion of atoms within the layer and in the orthogonal direction along the $c$-axis. Note that strong anisotropy of atom thermal vibrations has been also observed previously by single-crystal X-ray diffraction in 2H- and 3R-MoS$_2$ polytype phases \cite{Schonfeld1983}.  

\section{Conclusions}

Interlayer and intralayer coupling in two-dimensional layered 2H$_c$-MoS$_2$ was studied by 
the Mo K-edge X-ray absorption spectroscopy and ab initio molecular dynamics (AIMD) simulations.
The Mo K-edge EXAFS spectra were measured in the temperature range of 10-300 K, and their
analysis  using the reverse Monte Carlo (RMC) simulations allowed us to distinguish 
intralayer and interlayer interactions between absorbing Mo and neighboring S/Mo atoms located
within the same  or nearby layers, respectively. 
Temperature dependence of the  mean-square relative displacement factors $\sigma^2$ was determined for the nearest ten coordination shells around molybdenum. 
The strength of the Mo--S and Mo--Mo interatomic interactions was evaluated in terms of the characteristic Einstein frequencies $\omega_E$ and the effective force constants $\kappa$. 

The RMC results obtained at low temperatures were complemented by high-temperature AIMD simulations.  The experimental Mo K-edge EXAFS spectrum measured at 300~K was employed  
to validate the results of the AIMD simulations, which were used to extend the studied temperature range up to 1200~K.  Good agreement between the experimental and AIMD calculated EXAFS spectra at 300~K as well  as  between the mean-square relative displacement factors $\sigma^2$ for Mo--S and Mo--Mo atom pairs obtained by RMC and AIMD methods proves the reliability of the AIMD simulations. 
 
The AIMD results demonstrate good sensitivity to the temperature dependence of the interatomic interactions and provide evidence of the reduction of correlation in motion between distant atoms. Moreover, both RMC and AIMD results suggest strongly anisotropic vibrational dynamics of atoms within the plane of the layers and in the orthogonal direction along the $c$-axis. The demonstrated methodology opens a new pathway for the investigation of interatomic interactions in layered materials.

\section*{CRediT authorship contribution statement}

\textbf{Inga Pudza}: Investigation, Visualization, Writing – original draft, review \& editing.
\textbf{Dmitry Bocharov}: Conceptualization, Investigation, Writing – original draft, review \& editing.
\textbf{Andris Anspoks}: Investigation.
\textbf{Matthias Krack}: Methodology, Investigation.
\textbf{Aleksandr Kalinko}: Investigation.
\textbf{Edmund Welter}: Resources.
\textbf{Alexei Kuzmin}: Conceptualization, Investigation, Methodology, Writing – original draft, review \& editing.

\section*{Declaration of Competing Interest}

The authors declare that they have no known competing financial interests or personal relationships that could have appeared to influence the work reported in this paper.

\section*{Acknowledgements}
D.B. and A.K. thank the support of the Latvian Council of Science project No. LZP-2020/1-0261.
The experiment at the PETRA III synchrotron was performed within proposal No. I-20170739 EC.
The synchrotron experiment has been supported by the project CALIPSOplus under the Grant Agreement 730872 from the EU Framework Programme for Research and Innovation HORIZON 2020.
We acknowledge that the results of AIMD simulations have been achieved using the DECI resource Beskow based in Sweden at KTH Royal Institute of Technology in Stockholm with support from the PRACE (DECI15 project WS2MoS2AIMD). The technical support of Lilit Axner is gratefully acknowledged.  
Institute of Solid State Physics, University of Latvia as the Center of Excellence has received funding from the European Union’s Horizon 2020 Framework Programme H2020-WIDESPREAD-01-2016-2017-TeamingPhase2 under grant agreement No. 739508, project CAMART2.

\section*{Data availability statement}

Data will be made available on request.

\newpage

% \bibliography{mos2}

\begin{thebibliography}{10}
\expandafter\ifx\csname url\endcsname\relax
  \def\url#1{\texttt{#1}}\fi
\expandafter\ifx\csname urlprefix\endcsname\relax\def\urlprefix{URL }\fi
\expandafter\ifx\csname href\endcsname\relax
  \def\href#1#2{#2} \def\path#1{#1}\fi

\bibitem{Gupta2015}
A.~Gupta, T.~Sakthivel, S.~Seal, {Recent development in 2D materials beyond
  graphene}, Prog. Mater. Sci. 73 (2015) 44--126.
\newblock \href {https://doi.org/10.1016/j.pmatsci.2015.02.002}
  {\path{doi:10.1016/j.pmatsci.2015.02.002}}.

\bibitem{Sangwan2022}
V.~K. Sangwan, S.~E. Liu, A.~R. Trivedi, M.~C. Hersam, {Two-dimensional
  materials for bio-realistic neuronal computing networks}, Matter 5 (2022)
  4133--4152.
\newblock \href {https://doi.org/10.1016/j.matt.2022.10.017}
  {\path{doi:10.1016/j.matt.2022.10.017}}.

\bibitem{Novoselov2016}
K.~S. Novoselov, A.~Mishchenko, A.~Carvalho, A.~H. Castro~Neto, {2D materials
  and van der Waals heterostructures}, Science 353 (2016) aac9439.
\newblock \href {https://doi.org/10.1126/science.aac9439}
  {\path{doi:10.1126/science.aac9439}}.

\bibitem{Kumar2016}
P.~Kumar, H.~Abuhimd, W.~Wahyudi, M.~Li, J.~Ming, L.-J. Li,
  {Review—two-dimensional layered materials for energy storage applications},
  ECS J. Solid State Sci. Technol. 5 (2016) Q3021.
\newblock \href {https://doi.org/10.1149/2.0051611jss}
  {\path{doi:10.1149/2.0051611jss}}.

\bibitem{Zhan2019}
H.~Zhan, D.~Guo, G.~Xie, {Two-dimensional layered materials: from mechanical
  and coupling properties towards applications in electronics}, Nanoscale 11
  (2019) 13181--13212.
\newblock \href {https://doi.org/10.1039/C9NR03611C}
  {\path{doi:10.1039/C9NR03611C}}.

\bibitem{Gutierrez2020}
H.~R. Guti\'errez, {Two-dimensional layered materials offering expanded
  applications in flatland}, ACS Appl. Nano Mater. 3 (2020) 6134--6139.
\newblock \href {https://doi.org/10.1021/acsanm.0c01763}
  {\path{doi:10.1021/acsanm.0c01763}}.

\bibitem{Kaushik2021}
S.~Kaushik, R.~Singh, {2D layered materials for ultraviolet photodetection: A
  review}, Adv. Opt. Mater. 9 (2021) 2002214.
\newblock \href {https://doi.org/10.1002/adom.202002214}
  {\path{doi:10.1002/adom.202002214}}.

\bibitem{Lanza2022}
M.~Lanza, I.~Radu, {Electronic circuits made of 2D materials}, Adv. Mater. 34
  (2022) 2207843.
\newblock \href {https://doi.org/10.1002/adma.202207843}
  {\path{doi:10.1002/adma.202207843}}.

\bibitem{Hoang2022}
A.~T. Hoang, L.~Hu, A.~K. Katiyar, J.-H. Ahn, {Two-dimensional layered
  materials and heterostructures for flexible electronics}, Matter 5 (2022)
  4116--4132.
\newblock \href {https://doi.org/10.1016/j.matt.2022.10.016}
  {\path{doi:10.1016/j.matt.2022.10.016}}.

\bibitem{Shanmugam2022}
V.~Shanmugam, R.~A. Mensah, K.~Babu, S.~Gawusu, A.~Chanda, Y.~Tu, R.~E.
  Neisiany, M.~Försth, G.~Sas, O.~Das, {A review of the synthesis, properties,
  and applications of 2D materials}, Part. Part. Syst. Charact. 39 (2022)
  2200031.
\newblock \href {https://doi.org/10.1002/ppsc.202200031}
  {\path{doi:10.1002/ppsc.202200031}}.

\bibitem{Liu2016}
Y.~Liu, N.~Weiss, X.~Duan, H.-C. Cheng, Y.~Huang, X.~Duan, {Van der Waals
  heterostructures and devices}, Nat. Rev. Mater. 1 (2016) 16042.
\newblock \href {https://doi.org/10.1038/natrevmats.2016.42}
  {\path{doi:10.1038/natrevmats.2016.42}}.

\bibitem{Shi2018}
Z.~Shi, X.~Wang, Y.~Sun, Y.~Li, L.~Zhang, {Interlayer coupling in
  two-dimensional semiconductor materials}, Semicond. Sci. Technol. 33 (2018)
  093001.
\newblock \href {https://doi.org/10.1088/1361-6641/aad6c3}
  {\path{doi:10.1088/1361-6641/aad6c3}}.

\bibitem{Bian2021}
R.~Bian, C.~Li, Q.~Liu, G.~Cao, Q.~Fu, P.~Meng, J.~Zhou, F.~Liu, Z.~Liu,
  {Recent progress in the synthesis of novel two-dimensional van der Waals
  materials}, Natl. Sci. Rev. 9 (2021) nwab164.
\newblock \href {https://doi.org/10.1093/nsr/nwab164}
  {\path{doi:10.1093/nsr/nwab164}}.

\bibitem{Pak2018}
Y.~Pak, W.~Park, S.~Mitra, A.~A. Sasikala~Devi, K.~Loganathan, Y.~Kumaresan,
  Y.~Kim, B.~Cho, G.-Y. Jung, M.~M. Hussain, I.~S. Roqan, {Enhanced performance
  of MoS$_2$ photodetectors by inserting an ALD-processed TiO$_2$ interlayer},
  Small 14 (2018) 1703176.
\newblock \href {https://doi.org/10.1002/smll.201703176}
  {\path{doi:10.1002/smll.201703176}}.

\bibitem{Lee2015}
G.-H. Lee, X.~Cui, Y.~D. Kim, G.~Arefe, X.~Zhang, C.-H. Lee, F.~Ye,
  K.~Watanabe, T.~Taniguchi, P.~Kim, J.~Hone, {Highly stable, dual-gated
  MoS$_2$ transistors encapsulated by hexagonal boron nitride with
  gate-controllable contact, resistance, and threshold voltage}, ACS Nano 9
  (2015) 7019--7026.
\newblock \href {https://doi.org/10.1021/acsnano.5b01341}
  {\path{doi:10.1021/acsnano.5b01341}}.

\bibitem{Li2019}
W.~Li, J.~Zhou, S.~Cai, Z.~Yu, J.~Zhang, N.~Fang, T.~Li, Y.~Wu, T.~Chen,
  X.~Xie, H.~Ma, K.~Yan, N.~Dai, X.~Wu, H.~Zhao, Z.~Wang, D.~He, L.~Pan,
  Y.~Shi, P.~Wang, W.~Chen, K.~Nagashio, X.~Duan, X.~Wang, {Uniform and
  ultrathin high-$\kappa$ gate dielectrics for two-dimensional electronic
  devices}, Nat. Electron. 2 (2019) 563--571.
\newblock \href {https://doi.org/10.1038/s41928-019-0334-y}
  {\path{doi:10.1038/s41928-019-0334-y}}.

\bibitem{Das2021}
S.~Das, A.~Sebastian, E.~Pop, C.~J. McClellan, A.~D. Franklin, T.~Grasser,
  T.~Knobloch, Y.~Illarionov, A.~V. Penumatcha, J.~Appenzeller, Z.~Chen,
  W.~Zhu, I.~Asselberghs, L.-J. Li, U.~E. Avci, N.~Bhat, T.~D. Anthopoulos,
  R.~Singh, {Transistors based on two-dimensional materials for future
  integrated circuits}, Nat. Electron. 4 (2021) 786--799.
\newblock \href {https://doi.org/10.1038/s41928-021-00670-1}
  {\path{doi:10.1038/s41928-021-00670-1}}.

\bibitem{Desai2016}
S.~B. Desai, S.~R. Madhvapathy, A.~B. Sachid, J.~P. Llinas, Q.~Wang, G.~H. Ahn,
  G.~Pitner, M.~J. Kim, J.~Bokor, C.~Hu, H.-S.~P. Wong, A.~Javey, {MoS$_2$
  transistors with 1-nanometer gate lengths}, Science 354~(6308) (2016)
  99--102.
\newblock \href {https://doi.org/10.1126/science.aah4698}
  {\path{doi:10.1126/science.aah4698}}.

\bibitem{Wachter2017}
S.~Wachter, D.~K. Polyushkin, O.~Bethge, T.~Mueller, {A microprocessor based on
  a two-dimensional semiconductor}, Nat. Commun. 8 (2017) 14948.
\newblock \href {https://doi.org/10.1038/ncomms14948}
  {\path{doi:10.1038/ncomms14948}}.

\bibitem{Migliato2020logic}
G.~Migliato~Marega, Y.~Zhao, A.~Avsar, Z.~Wang, M.~Tripathi, A.~Radenovic,
  A.~Kis, {Logic-in-memory based on an atomically thin semiconductor}, Nature
  587 (2020) 72--77.
\newblock \href {https://doi.org/10.1038/s41586-020-2861-0}
  {\path{doi:10.1038/s41586-020-2861-0}}.

\bibitem{Ferrari2015}
A.~C. Ferrari, F.~Bonaccorso, V.~Fal{'}ko, K.~S. Novoselov, S.~Roche,
  P.~Boggild, S.~Borini, F.~H.~L. Koppens, V.~Palermo, N.~Pugno, J.~A. Garrido,
  R.~Sordan, A.~Bianco, L.~Ballerini, M.~Prato, E.~Lidorikis, J.~Kivioja,
  C.~Marinelli, T.~Ryhanen, A.~Morpurgo, J.~N. Coleman, V.~Nicolosi,
  L.~Colombo, A.~Fert, M.~Garcia-Hernandez, A.~Bachtold, G.~F. Schneider,
  F.~Guinea, C.~Dekker, M.~Barbone, Z.~Sun, C.~Galiotis, A.~N. Grigorenko,
  G.~Konstantatos, A.~Kis, M.~Katsnelson, L.~Vandersypen, A.~Loiseau,
  V.~Morandi, D.~Neumaier, E.~Treossi, V.~Pellegrini, M.~Polini, A.~Tredicucci,
  G.~M. Williams, B.~Hee~Hong, J.-H. Ahn, J.~Min~Kim, H.~Zirath, B.~J. van
  Wees, H.~van~der Zant, L.~Occhipinti, A.~Di~Matteo, I.~A. Kinloch,
  T.~Seyller, E.~Quesnel, X.~Feng, K.~Teo, N.~Rupesinghe, P.~Hakonen, S.~R.~T.
  Neil, Q.~Tannock, T.~Lofwander, J.~Kinaret, {Science and technology roadmap
  for graphene, related two-dimensional crystals, and hybrid systems},
  Nanoscale 7 (2015) 4598--4810.
\newblock \href {https://doi.org/10.1039/C4NR01600A}
  {\path{doi:10.1039/C4NR01600A}}.

\bibitem{Li2017}
X.~Li, L.~Tao, Z.~Chen, H.~Fang, X.~Li, X.~Wang, J.-B. Xu, H.~Zhu, {Graphene
  and related two-dimensional materials: Structure-property relationships for
  electronics and optoelectronics}, Appl. Phys. Rev. 4 (2017) 021306.
\newblock \href {https://doi.org/10.1063/1.4983646}
  {\path{doi:10.1063/1.4983646}}.

\bibitem{Butanovs2018}
E.~Butanovs, S.~Vlassov, A.~Kuzmin, S.~Piskunov, J.~Butikova, B.~Polyakov,
  {Fast-response single-nanowire photodetector based on ZnO/WS$_2$ core/shell
  heterostructures}, ACS Appl. Mater. Interfaces 10 (2018) 13869--13876.
\newblock \href {https://doi.org/10.1021/acsami.8b02241}
  {\path{doi:10.1021/acsami.8b02241}}.

\bibitem{Polyakov2016}
B.~Polyakov, A.~Kuzmin, K.~Smits, J.~Zideluns, E.~Butanovs, J.~Butikova,
  S.~Vlassov, S.~Piskunov, Y.~F. Zhukovskii, {Unexpected epitaxial growth of a
  few WS$_2$ layers on $\{1\bar{1}00\}$ facets of ZnO nanowires}, J. Phys.
  Chem. C 120 (2016) 21451--21459.
\newblock \href {https://doi.org/10.1021/acs.jpcc.6b06139}
  {\path{doi:10.1021/acs.jpcc.6b06139}}.

\bibitem{Billinge2007}
S.~J.~L. Billinge, I.~Levin, {The problem with determining atomic structure at
  the nanoscale}, Science 316 (2005) 561--565.
\newblock \href {https://doi.org/10.1126/science.1135080}
  {\path{doi:10.1126/science.1135080}}.

\bibitem{Stohr2019}
M.~St\"{o}hr, T.~Van~Voorhis, A.~Tkatchenko, {Theory and practice of modeling
  van der Waals interactions in electronic-structure calculations}, Chem. Soc.
  Rev. 48 (2019) 4118--4154.
\newblock \href {https://doi.org/10.1039/C9CS00060G}
  {\path{doi:10.1039/C9CS00060G}}.

\bibitem{Chung2014}
D.~Y. Chung, S.-K. Park, Y.-H. Chung, S.-H. Yu, D.-H. Lim, N.~Jung, H.~C. Ham,
  H.-Y. Park, Y.~Piao, S.~J. Yoo, Y.-E. Sung, {Edge-exposed MoS$_2$
  nano-assembled structures as efficient electrocatalysts for hydrogen
  evolution reaction}, Nanoscale 6 (2014) 2131--2136.
\newblock \href {https://doi.org/10.1039/C3NR05228A}
  {\path{doi:10.1039/C3NR05228A}}.

\bibitem{Caramazza2016}
S.~Caramazza, C.~Marini, L.~Simonelli, P.~Dore, P.~Postorino, {Temperature
  dependent EXAFS study on transition metal dichalcogenides MoX$_2$ (X = S, Se,
  Te)}, J. Phys.: Condens. Matter 28 (2016) 325401.
\newblock \href {https://doi.org/10.1088/0953-8984/28/32/325401}
  {\path{doi:10.1088/0953-8984/28/32/325401}}.

\bibitem{Kuznetsov2019}
D.~A. Kuznetsov, Z.~Chen, P.~V. Kumar, A.~Tsoukalou, A.~Kierzkowska, P.~M.
  Abdala, O.~V. Safonova, A.~Fedorov, C.~R. Muller, {Single site cobalt
  substitution in 2D molybdenum carbide (MXene) enhances catalytic activity in
  the hydrogen evolution reaction}, J. Am. Chem. Soc. 141 (2019) 17809--17816.
\newblock \href {https://doi.org/10.1021/jacs.9b08897}
  {\path{doi:10.1021/jacs.9b08897}}.

\bibitem{Kuzmin2014}
A.~Kuzmin, J.~Chaboy, {EXAFS and XANES analysis of oxides at the nanoscale},
  IUCrJ 1 (2014) 571--589.
\newblock \href {https://doi.org/10.1107/S2052252514021101}
  {\path{doi:10.1107/S2052252514021101}}.

\bibitem{Kuzmin2020}
A.~Kuzmin, J.~Timoshenko, A.~Kalinko, I.~Jonane, A.~Anspoks, {Treatment of
  disorder effects in X-ray absorption spectra beyond the conventional
  approach}, Rad. Phys. Chem. 175 (2020) 108112.
\newblock \href {https://doi.org/10.1016/j.radphyschem.2018.12.032}
  {\path{doi:10.1016/j.radphyschem.2018.12.032}}.

\bibitem{Rehr2000}
J.~J. Rehr, R.~C. Albers, {Theoretical approaches to x-ray absorption fine
  structure}, Rev. Mod. Phys. 72 (2000) 621--654.
\newblock \href {https://doi.org/10.1103/RevModPhys.72.621}
  {\path{doi:10.1103/RevModPhys.72.621}}.

\bibitem{Joensen1987}
P.~Joensen, E.~D. Crozier, N.~Alberding, R.~F. Frindt, {A study of single-layer
  and restacked MoS$_2$ by X-ray diffraction and X-ray absorption
  spectroscopy}, J. Phys. C: Solid State Physics 20 (1987) 4043.
\newblock \href {https://doi.org/10.1088/0022-3719/20/26/009}
  {\path{doi:10.1088/0022-3719/20/26/009}}.

\bibitem{BELYAKOVA2004}
O.~Belyakova, Y.~Zubavichus, I.~Neretin, A.~Golub, Y.~Novikov, E.~Mednikov,
  M.~Vargaftik, I.~Moiseev, Y.~Slovokhotov, {Atomic structure of nanomaterials:
  combined X-ray diffraction and EXAFS studies}, J. Alloys Compd. 382 (2004)
  46--53.
\newblock \href {https://doi.org/10.1016/j.jallcom.2004.05.047}
  {\path{doi:10.1016/j.jallcom.2004.05.047}}.

\bibitem{Lassalle2015}
B.~Lassalle-Kaiser, D.~Merki, H.~Vrubel, S.~Gul, V.~K. Yachandra, X.~Hu,
  J.~Yano, {Evidence from in situ X-ray absorption spectroscopy for the
  involvement of terminal disulfide in the reduction of protons by an amorphous
  molybdenum sulfide electrocatalyst}, J. Am. Chem. Soc. 137 (2015) 314--321.
\newblock \href {https://doi.org/10.1021/ja510328m}
  {\path{doi:10.1021/ja510328m}}.

\bibitem{Zhang2016}
H.~P. Zhang, H.~F. Lin, Y.~Zheng, Y.~F. Hu, A.~MacLennan, {The catalytic
  activity and chemical structure of nano MoS$_2$ synthesized in a controlled
  environment}, React. Chem. Eng. 1 (2016) 165--175.
\newblock \href {https://doi.org/10.1039/C5RE00046G}
  {\path{doi:10.1039/C5RE00046G}}.

\bibitem{KIM2017}
S.-H. Kim, K.-D. Kim, Y.-K. Lee, {Effects of dispersed MoS$_2$ catalysts and
  reaction conditions on slurry phase hydrocracking of vacuum residue}, J.
  Catal. 347 (2017) 127--137.
\newblock \href {https://doi.org/10.1016/j.jcat.2016.11.015}
  {\path{doi:10.1016/j.jcat.2016.11.015}}.

\bibitem{P65}
E.~Welter, R.~Chernikov, M.~Herrmann, R.~Nemausat, {A beamline for bulk sample
  x-ray absorption spectroscopy at the high brilliance storage ring PETRA III},
  AIP Conf. Proc. 2054 (2019) 040002.
\newblock \href {https://doi.org/10.1063/1.5084603}
  {\path{doi:10.1063/1.5084603}}.

\bibitem{Timoshenko2012rmc}
J.~Timoshenko, A.~Kuzmin, J.~Purans, {Reverse Monte Carlo modeling of thermal
  disorder in crystalline materials from EXAFS spectra}, Comp. Phys. Commun.
  183 (2012) 1237--1245.
\newblock \href {https://doi.org/10.1016/j.cpc.2012.02.002}
  {\path{doi:10.1016/j.cpc.2012.02.002}}.

\bibitem{Timoshenko2014rmc}
J.~Timoshenko, A.~Kuzmin, J.~Purans, {EXAFS study of hydrogen intercalation
  into ReO$_3$ using the evolutionary algorithm}, J. Phys.: Condens. Matter 26
  (2014) 055401.
\newblock \href {https://doi.org/10.1088/0953-8984/26/5/055401}
  {\path{doi:10.1088/0953-8984/26/5/055401}}.

\bibitem{Wyckoff1963}
R.~W.~G. Wyckoff, {Crystal Structures}, Vol.~1, Interscience Publishers, New
  York, 1963.

\bibitem{Ankudinov1998}
A.~L. Ankudinov, B.~Ravel, J.~J. Rehr, S.~D. Conradson, {Real-space
  multiple-scattering calculation and interpretation of x-ray-absorption
  near-edge structure}, Phys. Rev. B 58 (1998) 7565--7576.
\newblock \href {https://doi.org/10.1103/PhysRevB.58.7565}
  {\path{doi:10.1103/PhysRevB.58.7565}}.

\bibitem{Hedin1971}
L.~Hedin, B.~I. Lundqvist, {Explicit local exchange-correlation potentials}, J.
  Phys. C: Solid State Phys. 4 (1971) 2064.
\newblock \href {https://doi.org/10.1088/0022-3719/4/14/022}
  {\path{doi:10.1088/0022-3719/4/14/022}}.

\bibitem{Timoshenko2009wavelet}
J.~Timoshenko, A.~Kuzmin, {Wavelet data analysis of EXAFS spectra}, Comp. Phys.
  Commun. 180 (2009) 920--925.
\newblock \href {https://doi.org/10.1016/j.cpc.2008.12.020}
  {\path{doi:10.1016/j.cpc.2008.12.020}}.

\bibitem{Hampel1973robust}
F.~R. Hampel, {Robust estimation: A condensed partial survey}, Z.
  Wahrscheinlichkeit. 27 (1973) 87--104.
\newblock \href {https://doi.org/10.1007/BF00536619}
  {\path{doi:10.1007/BF00536619}}.

\bibitem{Daszykowski2007robust}
M.~Daszykowski, K.~Kaczmarek, Y.~Vander~Heyden, B.~Walczak, {Robust statistics
  in data analysis--A review: Basic concepts}, Chemom. Intell. Lab. Syst. 85
  (2007) 203--219.
\newblock \href {https://doi.org/10.1016/j.chemolab.2006.06.016}
  {\path{doi:10.1016/j.chemolab.2006.06.016}}.

\bibitem{Kohn1965}
W.~Kohn, L.~J. Sham, {Self-consistent equations including exchange and
  correlation effects}, Phys. Rev. 140 (1965) A1133--A1138.
\newblock \href {https://doi.org/10.1103/PhysRev.140.A1133}
  {\path{doi:10.1103/PhysRev.140.A1133}}.

\bibitem{CP2K2020}
T.~D. Kuhne, M.~Iannuzzi, M.~D. Ben, V.~V. Rybkin, P.~Seewald, F.~Stein,
  T.~Laino, R.~Z. Khaliullin, O.~Schutt, F.~Schiffmann, D.~Golze, J.~Wilhelm,
  S.~Chulkov, M.~H. Bani-Hashemian, V.~Weber, U.~Bor{\v{s}}tnik,
  M.~Taillefumier, A.~S. Jakobovits, A.~Lazzaro, H.~Pabst, T.~Muller,
  R.~Schade, M.~Guidon, S.~Andermatt, N.~Holmberg, G.~K. Schenter, A.~Hehn,
  A.~Bussy, F.~Belleflamme, G.~Tabacchi, A.~Glo{\ss}, M.~Lass, I.~Bethune,
  C.~J. Mundy, C.~Plessl, M.~Watkins, J.~VandeVondele, M.~Krack, J.~Hutter,
  {CP2K: An electronic structure and molecular dynamics software package --
  Quickstep: Efficient and accurate electronic structure calculations}, J.
  Chem. Phys. 152 (2020) 194103.
\newblock \href {https://doi.org/10.1063/5.0007045}
  {\path{doi:10.1063/5.0007045}}.

\bibitem{Quickstep}
J.~VandeVondele, M.~Krack, F.~Mohamed, M.~Parrinello, T.~Chassaing, J.~Hutter,
  {Quickstep: Fast and accurate density functional calculations using a mixed
  Gaussian and plane waves approach}, Comp. Phys. Commun. 167 (2005) 103--128.

\bibitem{GPW}
G.~Lippert, J.~Hutter, M.~Parrinello, A hybrid {Gaussian} and plane wave
  density functional scheme, Mol. Phys. 92 (1997) 477.

\bibitem{Perdew1996}
J.~P. Perdew, K.~Burke, M.~Ernzerhof, {Generalized gradient approximation made
  simple}, Phys. Rev. Lett. 77 (1996) 3865--3868.
\newblock \href {https://doi.org/10.1103/PhysRevLett.77.3865}
  {\path{doi:10.1103/PhysRevLett.77.3865}}.

\bibitem{Grimme2010}
S.~Grimme, J.~Antony, S.~Ehrlich, H.~Krieg, {A consistent and accurate ab
  initio parametrization of density functional dispersion correction (DFT-D)
  for the 94 elements H-Pu}, J. Chem. Phys. 132 (2010) 154104.
\newblock \href {https://doi.org/10.1063/1.3382344}
  {\path{doi:10.1063/1.3382344}}.

\bibitem{Bussi2007}
G.~Bussi, D.~Donadio, M.~Parrinello, {Canonical sampling through velocity
  rescaling}, J. Chem. Phys. 126 (2007) 014101.
\newblock \href {https://doi.org/10.1063/1.2408420}
  {\path{doi:10.1063/1.2408420}}.

\bibitem{Kuzmin2009}
A.~Kuzmin, R.~A. Evarestov, {Quantum mechanics--molecular dynamics approach to
  the interpretation of x-ray absorption spectra}, J. Phys.: Condens. Matter 21
  (2009) 055401.
\newblock \href {https://doi.org/10.1088/0953-8984/21/5/055401}
  {\path{doi:10.1088/0953-8984/21/5/055401}}.

\bibitem{Schonfeld1983}
B.~Sch{\"{o}}nfeld, J.~J. Huang, S.~C. Moss, {Anisotropic mean-square
  displacements (MSD) in single-crystals of 2H- and 3R-MoS$_2$}, Acta Cryst. B
  39 (1983) 404--407.
\newblock \href {https://doi.org/10.1107/S0108768183002645}
  {\path{doi:10.1107/S0108768183002645}}.

\bibitem{Jonane2018}
I.~Jonane, A.~Anspoks, A.~Kuzmin, {Advanced approach to the local structure
  reconstruction and theory validation on the example of the W L$_3$-edge
  extended x-ray absorption fine structure of tungsten}, Modelling Simul.
  Mater. Sci. Eng. 26 (2018) 025004.
\newblock \href {https://doi.org/10.1088/1361-651x/aa9bab}
  {\path{doi:10.1088/1361-651x/aa9bab}}.

\bibitem{Bakradze2021}
G.~Bakradze, A.~Kalinko, A.~Kuzmin, {Evidence of nickel ions dimerization in
  NiWO$_4$ and NiWO$_4$-ZnWO$_4$ solid solutions probed by EXAFS spectroscopy
  and reverse Monte Carlo simulations}, Acta Mater. 217 (2021) 117171.
\newblock \href {https://doi.org/10.1016/j.actamat.2021.117171}
  {\path{doi:10.1016/j.actamat.2021.117171}}.

\bibitem{Kotomin2022}
E.~A. Kotomin, A.~Kuzmin, J.~Purans, J.~Timoshenko, S.~Piskunov, R.~Merkle,
  J.~Maier, {Theoretical and experimental studies of charge ordering in
  CaFeO$_3$ and SrFeO$_3$ crystals}, Phys. Status Solidi B 259 (2022) 2100238.
\newblock \href {https://doi.org/10.1002/pssb.202100238}
  {\path{doi:10.1002/pssb.202100238}}.

\bibitem{Sevillano1979}
E.~Sevillano, H.~Meuth, J.~Rehr, {Extended x-ray absorption fine structure
  Debye-Waller factors. I. Monatomic crystals}, Phys. Rev. B 20 (1979) 4908.
\newblock \href {https://doi.org/10.1103/PhysRevB.20.4908}
  {\path{doi:10.1103/PhysRevB.20.4908}}.

\bibitem{Bagnall1980}
A.~Bagnall, W.~Liang, E.~Marseglia, B.~Welber, {Raman studies of MoS$_2$ at
  high pressure}, Physica B+C 99 (1980) 343--346.
\newblock \href {https://doi.org/10.1016/0378-4363(80)90257-0}
  {\path{doi:10.1016/0378-4363(80)90257-0}}.

\bibitem{Wakabayashi1975}
N.~Wakabayashi, H.~G. Smith, R.~M. Nicklow, {Lattice dynamics of hexagonal
  MoS$_2$ studied by neutron scattering}, Phys. Rev. B 12 (1975) 659--663.
\newblock \href {https://doi.org/10.1103/PhysRevB.12.659}
  {\path{doi:10.1103/PhysRevB.12.659}}.

\bibitem{Jeong2003}
I.-K. Jeong, R.~H. Heffner, M.~J. Graf, S.~J.~L. Billinge, {Lattice dynamics
  and correlated atomic motion from the atomic pair distribution function},
  Phys. Rev. B 67 (2003) 104301.
\newblock \href {https://doi.org/10.1103/PhysRevB.67.104301}
  {\path{doi:10.1103/PhysRevB.67.104301}}.

\end{thebibliography}

\end{document}